\documentclass[a4paper,11pt]{article} 
\pdfoutput=1 
\usepackage{jheppub} 
\usepackage{amsmath,graphicx,hyperref}

\newcommand{\fms}[1]{{#1}\!\!\!/}

\newcommand{\mc}{\mathcal}
\newcommand{\mr}{\mathrm}
\newcommand{\msf}{\mathsf}

\newcommand{\mO}{\mathcal{O}}

\newcommand{\be}{\begin{equation}} 
\newcommand{\ee}{\end{equation}} 
\newcommand{\bea}{\begin{eqnarray}} 
\newcommand{\eea}{\end{eqnarray}}

\newcommand{\pp}{\perp}

\newcommand{\n}{\overline{n}}
\newcommand{\nn}{\frac{\fms{\overline{n}}}{2}}

\newcommand{\blp}[1]{{\bf{#1}}_{\perp}}
\newcommand{\blpu}[1]{{\bf{#1}}^{\perp}}

\newcommand{\bsp}[1]{{\boldsymbol{#1}}_{\perp}}

\newcommand{\nnb}{\nonumber} 

\newcommand{\as}{\alpha_s}

\newcommand{\veps}{\varepsilon}

\begin{document}



\title{Heavy Quark Jet Fragmentation}

\def\Seoultech{Institute of Convergence Fundamental Studies and School of Liberal Arts,\\ Seoul National University of Science and Technology, Seoul 01811, Korea}
\def\Pitt{Pittsburgh Particle Physics Astrophysics and Cosmology Center (PITT PACC) \\ Department of Physics and Astronomy, University of Pittsburgh, Pittsburgh, Pennsylvania 15260, USA}

\author[a]{Lin Dai}
\emailAdd{lid33@pitt.edu}
\affiliation[a]{\Pitt}
\author[a,b]{Chul Kim}
\emailAdd{chul@seoultech.ac.kr}
\affiliation[b]{\Seoultech} 
\author[a]{Adam K. Leibovich}
\emailAdd{akl2@pitt.edu}

\abstract{
In this paper we study the fragmentation of a parton into a jet containing a heavy quark.
When heavy quarks are involved in a jet, the quark mass can lead to a numerically significant correction to the jet cross section and its substructure. 
With this motivation, we calculated the heavy quark mass effects to next-to-leading order in $\as$ on the fragmentation functions to a jet (FFJs) and the jet fragmentation functions (JFFs), where the former describes fragmentation of parton into a jet and the latter describes fragmenting processes inside a jet. 
The finite size of the heavy quark mass does not change the ultraviolet behaviors, but it can give significant corrections to the finite contributions.  
When we take the zero mass limit, we find that the FFJs and the JFFs reproduce established results for massless partons.   
If we define the heavy quark jet as one that include at least one heavy (anti-)quark, 
the tagged heavy quark jet production is sensitive to the heavy quark mass and produces large logarithms of the mass. Taking advantage of the FFJs and JFFs, we formulate a factorization theorem for  heavy quark jet production in order to resum these large logarithms systematically. As an application, we study  inclusive $b$-jet production and show phenomenological implications due to keeping a non-zero quark mass. 
}

\maketitle 


\section{Introduction} 

A jet, loosely defined to be a collimated beam of hadrons produced in a high energy collision, is well localized in a certain spatial direction and hence rather easy to experimentally identify. These objects are ubiquitous in high energy collisions.  A jet algorithm is used to map the momenta of the particles measured in the collision into a set of jets in a precise way.
To be theoretically useful, we only use jet algorithms that are infrared (IR) safe. This allows us to properly compare  theoretical results with  experiments, drastically reducing hadronic uncertainty. These are some of the reasons jet physics has become a crucial tool to test Standard Model and to unveil new physics.   

Interactions of particles related to a jet typically are offshell by an amount $p^2 \sim Q^2R^2$, where $Q$ is the typical hard energy scale and $R$ is a jet radius. 
So the appropriate scale that describes jet phenomena in hadron collisions is $\mu \sim p_TR$, where 
$p_T$ is the jet transverse momentum to an initial beam and is comparable to $Q$ for most of the rapidity region. 
When $R$ is an order of unity, the jet scale is similar to the hard collision scale of the initial partons, $\hat{s}^{1/2} \sim Q$. So, in this case, we have to describe the hard collision and the jet phenomena simultaneously. However, when $R$ is enough small, we can separate the hard collision and the jet processes properly and describe the jet phenomena using collinear interactions. 
Moreover, jets with a small radius are widely studied since they mitigate unwanted uncertainties arising from pile-up and underlying events.

Theoretically, a small jet radius is interesting because we can employ a collinear factorization theorem to write the rate as a convolution~\cite{Dasgupta:2014yra}
\be
\label{cofac} 
\frac{d\sigma (N_1N_2\to JX)}{dp_T^J} = \sum_{i}\int^1_{x_J} \frac{dz}{z}  \frac{d\sigma_{N_1N_2\to iX}}{dp_T^i}\Bigl(\frac{x_J}{z},\mu\Bigr) D_{J/i}(z,\mu)_,
\ee
where $i$ denotes a parton from the hard collision, and $p_T$ is a transverse momentum to an initial beam. $x_J$ is defined by $x_J=p_T^J/Q_T$, where $Q_T$ is the maximal transverse momentum of the parton $i$. $D_{J/i}(z)$ is the so-called fragmentation function to a jet (FFJ)~\cite{Kaufmann:2015hma,Kang:2016mcy,Dai:2016hzf}, 
which describes the probability for a mother parton $i$ to split into an observed jet $J$ with the transverse momentum fraction $z$. The factorization theorem in Eq.~(\ref{cofac}) encodes the fact that all the information on hard interactions at scale $\mu \sim p_T$ resides in the cross section $d\sigma/dp_T^i$, while the jet is properly described at the lower scale $\mu \sim p_TR$ described in terms of collinear interactions, where $R$ is the small jet radius. Thus the physical properties of the jet with a small $R$ is independent of the hard interactions and can be described completely by the FFJs.    

The FFJ in Eq.~(\ref{cofac}) acts like a plug-and-play module. If, instead of the FFJ, we insert a fragmentation function (FF) to a hadron, Eq.~(\ref{cofac}) describes a hadron $p_T$ spectrum. Due to this, the FFJs share many common features with the usual FFs. For example, the renormalization group (RG) running of the FFJs follows the well-known Dokshitzer-Gribov-Lipatov-Altarelli-Parisi (DGLAP) evolution as do the usual FFs. However, the low energy behavior of the FFJs is very different from a FF.  The FFJs are IR safe due to the finite size of the jet radius $R$. This fact enables us to estimate the FFJs by doing perturbative calculations including the resummation of the large logarithms of $R$~\cite{Dasgupta:2014yra,Kaufmann:2015hma,Kang:2016mcy,Dai:2016hzf,Dasgupta:2016bnd} and $1-z$~\cite{Dai:2017dpc}. Using these results, FFJs have been successfully used to calculate {\it inclusive} jet production~\cite{Kang:2016mcy,Dasgupta:2016bnd,Liu:2017pbb,Liu:2018ktv}.

The FFJs provide a firm basis to systematically explore the substructures of an observed jet. For example, when considering the fragmenting processes inside a jet, the scattering cross sections can be formulated as the multiplication of Eq.~(\ref{cofac}) and the jet fragmentation function (JFF)~\cite{Dai:2016hzf}, where the JFF describes the fragmentation within a jet to a particular hadron or a subjet, and has been widely studied~\cite{Kaufmann:2015hma,Dai:2016hzf,Procura:2011aq,Baumgart:2014upa,Chien:2015ctp,Kang:2016ehg,Bain:2017wvk,Kang:2017mda,Dai:2017cjq} on the basis of the theoretical results for the fragmenting jet functions~\cite{Procura:2009vm,Jain:2011xz,Ritzmann:2014mka}. Also, through the FFJs, we can consider the mass distribution~\cite{Idilbi:2016hoa,Kang:2018qra,Kang:2018jwa} and the transverse momentum distribution (to a jet axis)~\cite{Bain:2016rrv,Neill:2016vbi} for a jet with a given $p_T$.  

So far the FFJs and JFFs have been studied in using massless quarks only. 
It is therefore interesting to see how the quark mass affects the physical features when a heavy quark is in a jet.
If we consider a heavy quark jet at the LHC or a future collider using the factorization framework of Eq.~(\ref{cofac}), the quark mass can be safely ignored in the partonic cross section $d\sigma/dp_T^i$, since $p_T$ will be much larger than the heavy quark mass $m_Q$ in most cases, and so the mass can be set to zero in this part of the factorized formula. 
However, for a jet  with typical size $p_TR$, the quark mass may  be similar to this jet energy scale and can have a significant impact on the FFJ. 
In the limit $p_TR \sim m_Q$, the heavy quark can give a correction of order unity, and even in the limit $p_TR\gg m_Q$, the few first corrections in $m_Q/(p_TR)$ may be sizable. 

The heavy quark mass effects on the jet can be systematically studied in soft-collinear effective theory (SCET)~\cite{Bauer:2000ew,Bauer:2000yr,Bauer:2001yt,Bauer:2002nz}, which can be extended to include the quark mass~\cite{Leibovich:2003jd,Rothstein:2003wh,Chay:2005ck}. Using SCET, the jet can be described by collinear interactions with fluctuations $p_c^2 \sim p_T^2R^2$. As shown explicitly in Ref.~\cite{Chay:2005ck}, the massive version of SCET ($\mr{SCET_M}$) is renormalizable like full QCD. There it is shown that nontrivial ultraviolet (UV) effects due to the quark mass, like what occurs in heavy quark effective theory, do not appear in $\mr{SCET_M}$. Furthermore, it implies that the UV behavior of the heavy quark jet will be the same as for the massless case. We thus expect that the heavy quark mass can only change jet substructures or the low energy behavior of the jet. 

A heavy quark jet is usually defined by the inclusion of at least one heavy (anti-)quark.
However, as pointed out in Refs.~\cite{Banfi:2006hf,Banfi:2007gu}, if we specify the jet with a certain flavor of quark, we face large logarithms due to a hierarchy between the jet energy and the quark mass in the perturbative calculation. These logarithms become IR divergences in the massless limit.
If we consider $b$-jet production at the LHC in the regime $p_T \gg p_TR \gg m_b$, we would have the large logarithm $\ln p_TR/m_b$ at next-to-leading order (NLO) in $\as$. This logarithm arises when a gluon initiating a jet splits into a $b\bar{b}$-pair with a small opening angle. The logarithm would cancel if we consider the $b$-quark loop in the self-energy diagram of the gluon. However, this gluon self-energy diagram does not lead to $b$-quarks in the jet, and thus should not be included in the contribution to the $b$-jet.  Hence the large logarithm remains. 

In order to describe $b$-jet production and resum the large logarithms of $p_TR/m_b$, the factorization theorem in Eq.~(\ref{cofac}) alone is not enough. It is necessary to employ the JFF to a heavy quark while considering the substructure related to $g\to b\bar{b}$ further. The proper description of $b$-jet production can be realized through a factorization theorem with an appropriate combination of the FFJs and JFFs. For $p_T \gg p_TR \gg m_b$, the JFFs are responsible for the resummation of the large logarithms of $p_TR/m_b$, while the FFJs play an important role in resumming the logarithms of $R$. We further notice that the heavy quark JFF can be additionally factorized and matched onto the heavy quark fragmentation function (HQFF)~\cite{Mele:1990cw}. Then the large logarithm of $p_TR/m_b$ can be automatically resummed through RG running of the HQFF from $p_TR$ to $m_b$.\footnote{\baselineskip 3.0ex  A similar approach based on the calculation of the heavy quark fragmenting jet function in Ref.~\cite{Bauer:2013bza} has been considered in the context of the multi-jet production with a small $N$-jettiness. 
}

In this paper we study the heavy quark jet fragmenting processes and analyze the quark mass effects on the FFJs and JFFs. 
In Sec.~\ref{HQFFJsec} we extend the FFJs to include the heavy quark and calculate the heavy quark mass effects at NLO in $\as$. 
In Sec.~\ref{HQFFJ} we compute the heavy quark mass effects on the JFFs and confirm the established factorization formalism with the FFJs and JFFs up to NLO in $\as$. In Sec.~\ref{IbJp} we study  inclusive $b$-jet production at the LHC as an application. We show some phenomenological results using a  factorization theorem for $b$-jet production and resumming  large logarithms. In Sec.~\ref{concl} we  conclude.

\section{Next-to-leading order result of the FFJs with heavy quarks}
\label{HQFFJsec}

In order to effectively calculate the FFJs with heavy quarks at NLO in $\as$, it is useful to consider the one-loop computation of the HQFF at the parton level. The virtual one-loop correction to the HQFF automatically becomes the `in-jet' contribution and the real radiation can be separated into the `in-jet' and `out-jet' contributions by the jet algorithm. 

Following the definitions introduced in Ref.~\cite{Collins:1981uw} and using $\mr{SCET_M}$, we express the heavy quark and gluon fragmentation functions in $D$ dimensions as 
\bea
D_{i/Q}(z,\mu) &=& \sum_{X}\frac{1}{2N_cz}  \int d^{D-2}\blpu{p}_i  \mr{Tr} \langle 0 | \delta \Bigl(\frac{p_i^+}{z}-\mc{P}_+\Bigr)\delta^{(D-2)}(\bsp{\mc{P}})\nn \Psi_n^Q | i(p_i^+, \blpu{p}_i) X\rangle \nnb\\ 
\label{defHQFF1} 
&&\times\langle i(p_i^+, \blpu{p}_i) X | \bar{\Psi}_n^Q |0\rangle,\\
D_{i/g} (z,\mu) &=&  \sum_{X} \frac{1}{p_i^+(D-2)(N_c^2-1)} \int d^{D-2}\blpu{p}_i 
\mr{Tr} \langle 0 |  \delta\Bigl(\frac{p_i^+}{z}-\mc{P}_+\Bigr)\delta^{(D-2)}(\bsp{\mc{P}})  \mc{B}_n^{\pp\mu,a}\nnb \\
\label{defgFF1}
&&\times | i(p_i^+, \blpu{p}_i) X\rangle  \langle i(p_i^+, \blpu{p}_i) X | \mc{B}_{n\mu}^{\pp a}| 0 \rangle_, 
\eea 
where $i=q,Q,g$ denote flavors of partons, $q~(Q)$ is a light (heavy) quark, and $N_c$ is the number of colors. The fragmentation functions defined here are written in terms of collinear fields in the $n$ direction. For $n$-collinear interactions, the momentum is power counted as $p^{\mu} = (p_+,p_{\perp},p_-) = p_+(1,\lambda,\lambda^2)$, where $p_+\equiv \n\cdot p$, $p_-\equiv n\cdot p$,  and $\lambda$ is a small parameter dependent on the kinematic situation. We are using the standard lightcone vectors $n$ and $\n$ with normalization $n\cdot \n=2$ and gauge invariant collinear quark field and gluon field strength, 
$\Psi_n = W_n^{\dagger} \xi_n$ and $\mc{B}_{n}^{\pp\mu,a}  = i\n^{\rho}g_{\perp}^{\mu\nu} G_{n,\rho\nu}^b \mc{W}_n^{ba} = i\n^{\rho}g_{\perp}^{\mu\nu} \mc{W}_n^{\dagger,ba} G_{n,\rho\nu}^b$, respectively. Finally, $W_n~(\mc{W}_n$) is a collinear Wilson line in the fundamental (adjoint) representation.   

In expressing Eqs.~(\ref{defHQFF1}) and (\ref{defgFF1}), we have set the transverse momentum of the mother parton, $\blp{q} = \blpu{p}_i + \blpu{p}_X$, to zero. It is also sometimes useful to work in the frame where the transverse momentum of the observed parton $\blpu{p}_i =0$. In this case, after a slight rotation, we can express the fragmentation functions as follows: 
\bea
\label{defHQFF2}
D_{i/Q}(z,\mu) &=& \sum_{X}\frac{z^{D-3}}{2N_c} \mr{Tr} \langle 0 | \delta \Bigl(\frac{p_i^+}{z}-\mc{P}_+\Bigr)\nn \Psi_n^Q | i(p_i^+) X\rangle \langle i(p_i^+) X | \bar{\Psi}_n^Q |0\rangle,\\
\label{defgFF2}
D_{i/g} (z,\mu) &=&  \sum_{X} \frac{z^{D-2}}{p_i^+(D-2)(N_c^2-1)} 
\mr{Tr} \langle 0 |  \delta\Bigl(\frac{p_i^+}{z}-\mc{P}_+\Bigr) \mc{B}_n^{\pp\mu,a}| i(p_i^+) X\rangle \\
&&\hspace{4cm}\times \langle i(p_i^+) X | \mc{B}_{n\mu}^{\pp a}| 0 \rangle_. \nnb
\eea
 
In order to calculate at one-loop order, we will use an inclusive $\mr{k_T}$-type jet algorithm to include $\mr{k_T}$~\cite{Catani:1993hr,Ellis:1993tq}, C/A~\cite{Dokshitzer:1997in}, and anti-$\mr{k_T}$~\cite{Cacciari:2008gp} jets. When two emitted particles are combined into a jet, the constraint is given by 
\be
\label{jmerging} 
\theta < R',
\ee 
where $\theta$ is the angle between the two particles. $R'=R$ for $e^+e^-$ annihilation and $R'=R/\cosh{y}$ for hadron collision, where $R$ is the jet radius and $y$ is the rapidity. We will assume that $R$ is small enough to describe a jet using collinear interactions and $|y| \lesssim \mO(1)$ to constrain the event to the central region of the detector. Therefore, if we have a jet with energy $E_J$, the typical scale for describing the jet can be chosen to be $\mu\sim E_JR'$, which is given by $E_JR$ for $e^+e^-$ annihilations and $p_T^J R$ for hadron collisions, with $p_T^J$ being the transverse momentum of the jet relative to the beam direction. 

From Eq.~(\ref{jmerging}), we obtain the phase space constraint for jet merging when we have a splitting $q\to p+k$, where $q$ is a momentum of the mother parton and $p$ is the momentum of the observed parton,
\bea
\label{pconq}
\tan^2 \Bigl(\frac{R'}{2}\Bigr) &>& \frac{q_+^2\blp{k}^2}{p_+^2k_+^2},~~~~~~(\blp{q} = 0),\\
\label{pconp}
\tan^2 \Bigl(\frac{R'}{2}\Bigr) &>& \frac{\blp{k}^2}{k_+^2},~~~~~~~~~(\blp{p} = 0).
\eea
These constraints hold for both massless and massive partons, as long as the particles' masses are much smaller than their energies.  

Throughout this paper, we will renormalize using dimensional regularization with $D=4-2\veps$ and use the $\mr{\overline{MS}}$ scheme. In regularizing, we do not separate ultraviolet (UV) and infrared (IR) divergences for convenience, because the divergence structures for jet fragmentation have already been understood from the massless calculation. (For details, we refer to Ref.~\cite{Dai:2016hzf}.) Some differences occur in IR poles when comparing the massive and massless cases. As is seen in the HQFF, some IR poles in the massless case will be replaced with the logarithms with the quark mass $\ln (\mu^2/m^2)$.

\subsection{Heavy Quark Initiated Processes}

\begin{figure}[t]
\begin{center}
\includegraphics[height=5.5cm]{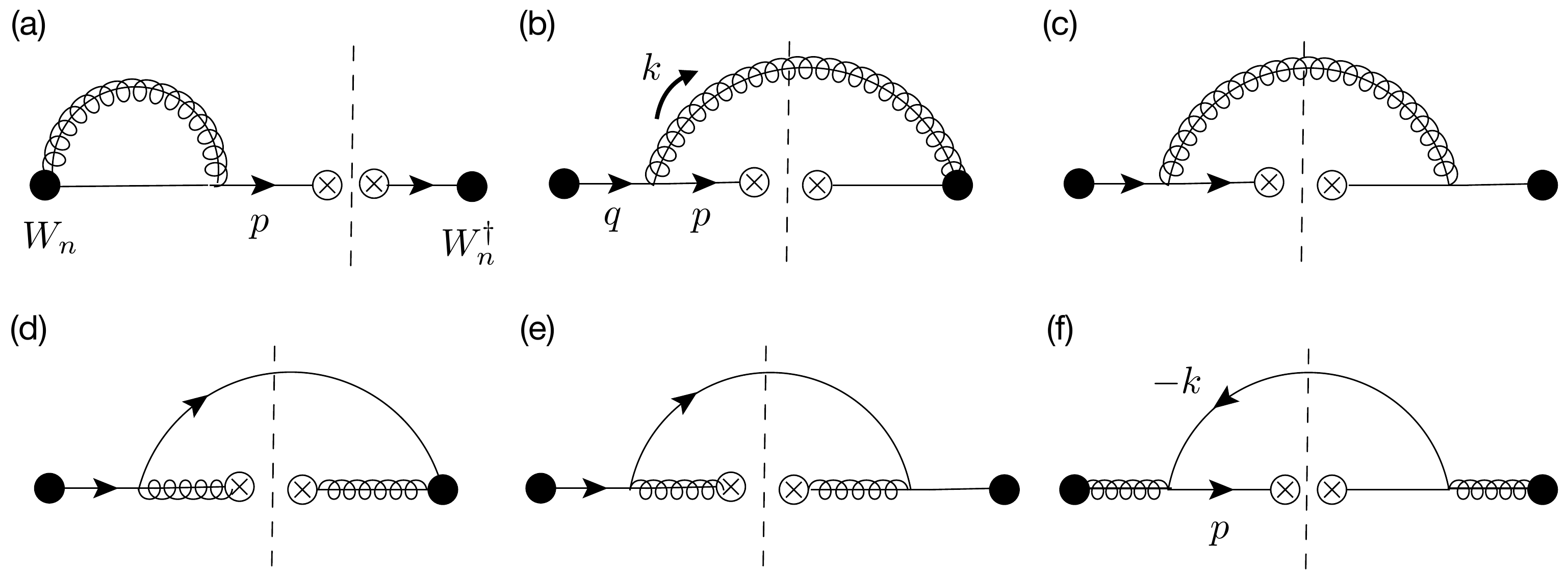}
\end{center}
\vspace{-0.5cm}
\caption{\label{fig1}
Fragmenting processes related to a heavy quark ($Q$). The dashed lines denote final state cuts. Diagrams (a), (b), and (c) describes the $Q\to Q$ process, while  diagrams (d) and (e) are $Q\to g$. Diagram (f) represents the $g\to Q$ process. Diagrams (a), (b), and (d)  each have a mirror diagram not shown. }
\end{figure}

In this subsection, we compute the NLO corrections to the fragmenting processes initiated by a heavy quark described in Eq.~(\ref{defHQFF1}) or (\ref{defHQFF2}). We first consider the $Q\to Q$ processes. The Feynman diagrams for the virtual and real contributions at NLO are shown in Fig.~\ref{fig1}-(a-c). The virtual contributions, arising from calculating Fig.~\ref{fig1}-(a) and its mirror diagram, are 
\be
\label{QQv} 
\mc{M}^V_{Q\to Q} (z;m,\mu) =\delta(1-z)  \frac{\as C_F}{2\pi}  \Bigl[\frac{1}{\veps^2}+\frac{1}{\veps}\Bigl(2+\ln\frac{\mu^2}{m^2}\Bigr) 
+2\ln\frac{\mu^2}{m^2}+\frac{1}{2} \ln^2\frac{\mu^2}{m^2}+4+\frac{\pi^2}{12}\Bigr],
\ee
where $m$ is the heavy quark mass. 

The real contributions from the diagrams Fig.~\ref{fig1}-(b) (and its mirror) and Fig.~\ref{fig1}-(c) are divided into in-jet and out-jet contributions through the jet algorithm in Eq.~(\ref{jmerging}). The in-jet real contributions are  
\begin{align}
\mc{M}^{R,\mr{In}}_{Q\to Q} (z;q_+t,m,\mu) &= \frac{\as C_F}{2\pi} \Biggl\{\delta(1-z) \Biggl[-\Bigl(\frac{1}{\veps}+\ln\frac{\mu^2}{q_+^2t^2}+\frac{1}{2}\Bigr) \ln\frac{q_+^2t^2+m^2}{m^2}-\frac{\pi^2}{6}-\frac{1}{2}\ln^2\frac{m^2}{q_+^2 t^2} \nnb \\
&+\frac{q_+^2t^2}{q_+^2t^2+m^2}\Bigl(\frac{1}{\veps}+\ln\frac{\mu^2}{q_+^2t^2}+2\Bigr)- {\rm Li}_2 \bigl(-\frac{m^2}{q_+^2 t^2}\bigr)+f\bigl(\frac{m^2}{q_+^2 t^2}\bigr)+g\bigl(\frac{m^2}{q_+^2 t^2}\bigr)\Biggr]\nnb \\
\label{QQRin} 
&+\Biggl[\frac{1+z^2}{1-z^2}\ln \frac{z^2 q_+^2 t^2+m^2}{m^2}-\frac{2z}{1-z} \frac{z^2 q_+^2t^2}{z^2 q_+^2t^2+m^2}\Biggr]_{+}\Biggr\},
\end{align}
where $t\equiv \tan (R'/2)$, $z=p_+/q_+$, and $q~(p)$ is the momentum of the mother (observed) parton. The subscript `+' in the brackets denotes the standard plus function. In the part proportional to $\delta (1-z)$ of Eq.~(\ref{QQRin}), the functions $f$ and $g$ are the following integrals
\bea
\label{intf}
f(b) &=& \int^1_0 dx \frac{1+x^2}{1-x} \ln \frac{x^2+b}{1+b}, \\
\label{intg}
g(b) &=& -2 \int^1_0 dx \frac{x}{1-x} \Bigl(\frac{x^2}{x^2+b}-\frac{1}{1+b}\Bigr).
\eea
In the limit of $b\to 0$, these functions becomes $f(0) = 5/2-2\pi^2/3$ and $g(0) = 0$.
We also computed the out-jet real contribution, with the result
\begin{align}
\mc{M}^{R,\mr{Out}}_{Q\to Q} (z;q_+t,m,\mu) &= \frac{\as C_F}{2\pi} \Biggl\{\delta(1-z) \Biggl[-\frac{1}{\veps^2}-\frac{1}{\veps}\Bigl(\ln\frac{\mu^2}{q_+^2t^2+m^2}+\frac{3}{2}\Bigr) -\frac{3}{2}\ln\frac{\mu^2}{q_+^2t^2+m^2}  \nnb \\
&-\frac{1}{2}\ln^2\frac{\mu^2}{q_+^2 t^2+m^2} -4+\frac{\pi^2}{12}+\frac{1}{2}\ln^2\frac{q_+^2 t^2+m^2}{q_+^2 t^2} -\ln\frac{q_+^2 t^2+m^2}{q_+^2 t^2} \nnb \\
& +\frac{m^2}{q_+^2t^2+m^2}\Bigl(\frac{1}{\veps}+\ln\frac{\mu^2}{q_+^2t^2}+2\Bigr)+ {\rm Li}_2 \bigl(-\frac{m^2}{q_+^2 t^2}\bigr)-f\bigl(\frac{m^2}{q_+^2 t^2}\bigr)-g\bigl(\frac{m^2}{q_+^2 t^2}\bigr)\Biggr]\nnb \\
&+\Biggl[\frac{1+z^2}{1-z^2}\Bigl(\frac{1}{\veps}+\ln \frac{\mu^2}{z^2 q_+^2 t^2+m^2}-2\ln(1-z)\Bigr)-(1-z)\nnb \\
\label{QQRout} 
&-\frac{2z}{1-z} \frac{m^2}{z^2 q_+^2t^2+m^2}\Biggr]_{+}\Biggr\}.
\end{align}

Combining the in-jet and out-jet contributions, the net result recovers the usual full splitting process for $Q\to Q$. We can thus reproduce the one-loop corrections to the HQFF as follows: 
\bea
D_{Q/Q}^{(1)}(z;m,\mu) &=& \mc{M}^V_{Q\to Q}+\bigl(\mc{M}^{R,\mr{In}}_{Q\to Q}+\mc{M}^{R,\mr{Out}}_{Q\to Q}\bigr)+\bigl(Z_{Q}^{(1)}+R_{Q}^{(1)}\bigr)\delta(1-z) \nnb \\
\label{DQQ}
&=&\frac{\as C_F}{2\pi} \Biggl[\frac{1+z^2}{1-z}\Bigl(\frac{1}{\veps}+\ln\frac{\mu^2}{m^2(1-z)^2}-1\Bigr)\Biggr]_{+},
\eea
where $Z_Q$ and $R_Q$ are the heavy quark renormalization and residue, respectively, with the one-loop expressions
\bea
\label{ZQ}
Z_Q^{(1)} &=& - \frac{\as C_F}{4\pi} \frac{1}{\veps},\\ 
\label{RQ}
R_Q^{(1)} &=& - \frac{\as C_F}{4\pi} \Bigl(\frac{2}{\veps} + 3\ln\frac{\mu^2}{m^2}+4\Bigr). 
\eea

Diagrams, Fig.~\ref{fig1}-(d) (and its mirror) and Fig.~\ref{fig1}-(e), contribute to the $Q\to g$ process. In-jet and out-jet contributions are, respectively,
\bea 
\mc{M}^{R,\mr{In}}_{Q\to g} (z;q_+t,m,\mu) &=& \frac{\as C_F}{2\pi} \Biggl[\frac{1+(1-z)^2}{z} \ln \frac{(1-z)^2 q_+^2 t^2+m^2}{m^2} \nnb \\
\label{Qgin}
&&~~~~ -2 \frac{1-z}{z} \frac{(1-z)^2 q_+^2 t^2}{(1-z)^2 q_+^2 t^2+m^2}\Biggr], \\
\mc{M}^{R,\mr{Out}}_{Q\to g} (z;q_+t,m,\mu) &=& \frac{\as C_F}{2\pi} \Biggl[\frac{1+(1-z)^2}{z} \Bigl(\frac{1}{\veps}+\ln\frac{\mu^2}{(1-z)^2 q_+^2 t^2+m^2}-2\ln z\Bigr)-z \nnb \\
\label{Qgout}
&&~~~~ -2 \frac{1-z}{z} \frac{m^2}{(1-z)^2 q_+^2 t^2+m^2}\Biggr].
\eea
Combining Eqs.~(\ref{Qgin}) and (\ref{Qgout}), we also reproduce the one-loop result of $Q\to g$ for the HQFF, 
\be
\label{DgQ}
D^{(1)}_{g/Q} (z;m,\mu) = \frac{\as C_F}{2\pi} \frac{1+(1-z)^2}{z} \Bigl(\frac{1}{\veps}+\ln\frac{\mu^2}{m^2(1-z)^2}-1\Bigr).
\ee

In order to describe inclusive radiations inside the jet initiated by a heavy quark, we introduce the heavy quark integrated jet function\footnote{The integrated jet function is also called `the unmeasured jet function', as introduced in Ref.~\cite{Ellis:2010rwa}},
\be
\label{HQIJ} 
\mc{J}_Q(E_JR',m,\mu) = \sum_{X_n \in J}\frac{1}{2N_c p_J^+} \mr{Tr}\langle 0 |\nn \Psi_n^Q |QX_n\in J(E_J,R')\rangle 
\langle QX_n\in J|\bar{\Psi}_n^Q | 0 \rangle_.
\ee
This is normalized to one at LO in $\as$. 
The heavy quark jet function describes events where the radiation off the initial heavy quark stays within the jet. It therefore can directly enter when we consider exclusive heavy quark jet cross sections. It also can be used as a normalization factor when we consider substructure distributions, including the JFF introduced in Eq.~\eqref{factJFF}. 
Applying the momentum sum rule to the results of Eqs.~(\ref{QQv}), (\ref{QQRin}), and (\ref{Qgin}), we can obtain the one-loop result of the heavy quark integrated jet function: 
\be
\label{mcsJ}
\mc{J}_Q^{(1)} = \int^1_0 dz z \Bigl[\mc{M}_{Q\to Q}^V (z) + \mc{M}_{Q\to Q}^{R,\mr{In}}(z)+\mc{M}_{Q\to g}^{R,\mr{In}}(z)\Bigr]
+Z_Q^{(1)}+R_Q^{(1)}.
\ee
As a result the renormalized heavy quark integrated jet function at NLO is 
\bea
\mc{J}_Q(E_JR',m,\mu) &=& 1+ \frac{\as C_F}{2\pi} \Biggl[\frac{1}{2} \ln\frac{\mu^2}{p_J^{+2}t^2+m^2}
+\frac{p_J^{+2} t^2}{p_J^{+2}t^2+m^2}\Bigl(\ln\frac{\mu^2}{p_J^{+2}t^2}+2\Bigr) +\frac{1}{2} \ln^2\frac{\mu^2}{p_J^{+2}t^2+m^2} 
\nnb \\
\label{NLOintQJ}
&&-\frac{1}{2}\ln^2\frac{p_J^{+2}t^2+m^2}{p_J^{+2}t^2}+2-\frac{\pi^2}{12}
-{\rm Li}_2 \bigl(-\frac{m^2}{p_J^{+2}t^2}\bigr)+f\bigl(\frac{m^2}{p_J^{+2}t^2}\bigr)+g\bigl(\frac{m^2}{p_J^{+2}t^2}\bigr)\Biggr] ,
\eea 
where $p_J^{+}t$ can be approximated as $E_JR'$. The result is IR finite. Moreover it does not involve the term $\ln(\mu^2/m^2)$, which represents the low energy dynamics with fluctuations of order $p^2\sim m^2$ if we consider the limit $E_J R' \gg m$. We also checked that, as $m$ goes to zero, $\mc{J}_Q(\mu;E_JR',m)$ becomes the same as the integrated jet function initiated by a light quark, of which the NLO results are~\cite{Ellis:2010rwa,Cheung:2009sg,Chay:2015ila} 
\be
\label{intqJ} 
\mc{J}_q(E_JR',\mu) = 1+ \frac{\as C_F}{2\pi} \Bigl[\frac{3}{2} \ln\frac{\mu^2}{p_J^{+2}t^2}
+\frac{1}{2}\ln^2\frac{\mu^2}{p_J^{+2}t^2} +\frac{13}{2} - \frac{3\pi^2}{4}\Bigr].
\ee

We now have all ingredients needed to compute the FFJ initiated by a heavy quark. At the operator level, it is defined as
\begin{align}
D_{J_i/Q}(z;ER',m,\mu) &= \sum_{X \notin J,X_{J-1}}\frac{1}{2N_cz}  \int d^{D-2}\blpu{p}_J  \mr{Tr} \langle 0 | \delta \Bigl(\frac{p_J^+}{z}-\mc{P}_+\Bigr)\delta^{(D-2)}(\bsp{\mc{P}})\nn \Psi_n^Q  \nnb\\ 
\label{defHQFFJ} 
&~~~~~~~~~~~\times| J_i(p_J^+, \blpu{p}_J) X_{\notin J}\rangle
\langle J_i(p_J^+, \blpu{p}_J) X_{\notin J} | \bar{\Psi}_n^Q |0\rangle,
\end{align} 
where $J_i$ is the jet initiated by the parton $i=Q,g$. $X_{J-1}$ are the final states in $J_i$, not including the primary parton $i$. $ER'$ in the argument of the heavy quark FFJ is an approximation of $q_+t$, where $q_+ = p_J^+/z$ is the large momentum component of the mother parton.
Up to NLO in $\as$, we find
\bea 
\label{indHQFFJQ}
D_{J_Q/Q} (z) &=&   \delta(1-z) \cdot\mc{J}_Q+ \mc{M}_{Q\to Q}^{R,\mr{Out}}(z), \\
\label{indHQFFJg}
D_{J_g/Q} (z) &=&   \mc{M}_{Q\to q}^{R,\mr{Out}}(z),
\eea 
where the one-loop results for $\mc{M}_{Q\to Q}^{R,\mr{Out}}$ and $\mc{M}_{Q\to g}^{R,\mr{Out}}$ are shown in Eqs.~(\ref{QQRout}) and (\ref{Qgout}), respectively. 

From Eqs.~(\ref{indHQFFJQ}) and (\ref{indHQFFJg}), the renormalized results are given by  
\bea
D_{J_Q/Q}(z;ER',m,\mu)  &=& \delta(1-z) + \frac{\as C_F}{2\pi}\Biggl\{\delta(1-z) \Bigl[f\bigl(\frac{m^2}{p_J^{+2} t^2}\bigr)+g'\bigl(\frac{m^2}{p_J^{+2} t^2}\bigr)\Bigr] \nnb \\
&&+\Bigl(\frac{1+z^2}{1-z}\Bigr)_+ \ln\frac{\mu^2}{z^2 q_+^2 t^2+m^2}-\Bigl(2\frac{1+z^2}{1-z}\ln(1-z)+1-z\Bigr)_+ \nnb\\
\label{DQQnlo}
&&-\Bigl(\frac{2z}{1-z}\Bigr)_+\frac{m^2}{z^2 q_+^2 t^2+m^2} \Biggr\},\\
D_{J_g/Q}(z;ER',m,\mu)  &=& \frac{\as C_F}{2\pi}\Biggl\{\frac{1+(1-z)^2}{z}\Bigl(\ln\frac{\mu^2}{(1-z)^2 q_+^2 t^2+m^2}-2\ln z\Bigr)-z\nnb\\
\label{DQgnlo}
&&-\frac{2(1-z)}{z} \frac{m^2}{(1-z)^2 q_+^2 t^2+m^2}\Biggr\}\ , 
\eea 
where $g'(b)$ in eq.~\eqref{DQQnlo} is given by 
\be 
g'(b) = 2 \int^1_0 dx \frac{x}{1-x} \Bigl(\frac{b}{x^2+b}-\frac{b}{1+b}\Bigr).
\ee
Here we expressed the FFJs in terms of $q_+t$ rather than $p_J^+t$ to manifestly show the momentum sum rule
\be
\label{sumrK} 
\sum_i \int^1_0 dz z D_{J_i/Q}(z) = 1.
\ee
If we rewrite the FFJs with $p_J^+t$ using $p_J^+=z q_+$, the sum rule does not hold. In obtaining Eq.~(\ref{DQQnlo}) from Eq.~(\ref{indHQFFJQ}), we found that the piece proportional to $\delta(1-z)$ in $\mc{M}_{Q\to Q}^{R,\mr{Out}}$ in Eq.~(\ref{QQRout}) is cancelled by the one-loop result of $\mc{J}_{Q}$ in Eq.~(\ref{NLOintQJ}). This results from the fact that the sum of the integrated jet function inside and outside the jet is given by 1 to all orders in $\as$. 

If we take the limit $m\to 0$ in Eqs.~(\ref{DQQnlo}) and (\ref{DQgnlo}), the results become the same as the light quark FFJs, $D_{J_q/q}$ and $D_{J_g/q}$, which are given in Ref.~\cite{Dai:2016hzf}. Furthermore, the renormalization group (RG) evolutions of the heavy quark FFJs follow DGLAP evolutions similar to the light quark FFJs, since the heavy quark mass does not affect the UV behavior. When we compute Eqs.~(\ref{DQQnlo}) and (\ref{DQgnlo}), we have assumed no scale hierarchy between $E_JR'$ and $m$. Thus, the results are valid in the limit $E_JR' \sim m$. 
Note that these results are also useful in the limit $E_JR' \gg m$. In this case, Eqs.~(\ref{DQQnlo}) and (\ref{DQgnlo}) can be understood as the resummed results to all orders in the mass correction such as $m^2/(E_JR')^2$.

\subsection{Gluon Initiated Processes}
\label{GIP}

Similar to the quark FFJ presented in Eq.~(\ref{defHQFFJ}), the gluon FFJs at the operator level is defined as 
\bea\label{gJFFpar}
D_{J_i/g} (z;ER',m_i,\mu) &=&  \sum_{X_{\notin J},X_{J-1}} \frac{1}{p_J^+(D-2)(N_c^2-1)} \int d^{D-2}\blpu{p}_J  \\
&&\hspace{-2.5cm}\times \mr{Tr} \langle 0 |  \delta\Bigl(\frac{p_J^+}{z}-\mc{P}_+\Bigr)\delta^{(D-2)}(\bsp{\mc{P}})  \mc{B}_n^{\pp\mu,a}
| J_i(p_J^+,\blpu{p}_J,R) X_{\notin J}\rangle 
\langle J_i(p_J^+,\blpu{p}_J,R) X_{\notin J} | \mc{B}_{n\mu}^{\pp a}| 0 \rangle_. \nnb
\eea
For these gluon initiated processes, the heavy-quark mass effect appears only from the heavy-quark loop diagram shown in Fig.~\ref{fig1}-(f). Starting from Eq.~(\ref{defgFF1}) or Eq.~(\ref{defgFF2}) with $i=Q$, the heavy quark loop contributions to the in-jet and out-jet are  
\bea
\label{MgQin}
\mc{M}_{g\to Q}^{R,\mr{In}} (z) &=& \frac{\as }{4\pi} \Biggl\{\Bigl[z^2+(1-z)^2\Bigr]
\ln \frac{z^2(1-z)^2 q_+^2t^2+m^2}{m^2} \\
&&~~~~~~+2z(1-z) \frac{z^2(1-z)^2 q_+^2t^2}{z^2(1-z)^2 q_+^2t^2+m^2}\Biggr\},\nnb \\
\label{MgQout}
\mc{M}_{g\to Q}^{R,\mr{Out}} (z) &=& \frac{\as}{4\pi} \Biggl\{\Bigl[z^2+(1-z)^2\Bigr]\Bigl(\frac{1}{\veps}
+ \ln \frac{\mu^2}{z^2(1-z)^2 q_+^2t^2+m^2} \Bigr)\\
&&~~~~~~-2z(1-z) \frac{z^2(1-z)^2 q_+^2t^2}{z^2(1-z)^2 q_+^2t^2+m^2}\Biggr\},\nnb
\eea
where $q_+$ is the momentum of the mother parton (gluon), and the $g\to \bar{Q}$ contributions are given by the same expressions. 

Including the in-jet contributions from $g\to g$ and $g\to q(\bar{q})$ processes and adding all the contributions as we did in  Eq.~(\ref{mcsJ}), we can obtain the integrated jet function (inside a jet) initiated by gluon. The renormalized result at NLO is given by
\bea
\mc{J}_g(E_JR',m_i,\mu) &=& 1+ \frac{\as C_A}{2\pi} \Biggl\{\frac{\beta_0}{2C_A} \ln\frac{\mu^2}{p_J^{+2}t^2}
+\frac{1}{2} \ln^2\frac{\mu^2}{p_J^{+2}t^2} +\frac{67}{9}-\frac{3\pi^2}{4} -\frac{23}{18}\frac{n_q}{C_A}
\nnb \\
\label{NLOintgJ}
&&\phantom{1+ \frac{\as C_A}{2\pi} \Biggl\{}
+\frac{1}{C_A}\sum_{i=1}^{n_Q} \Bigl[h\bigl(\frac{m_i^2}{p_J^{+2}t^2}\bigr) +j\bigl(\frac{m_i^2}{p_J^{+2}t^2}\bigr) \Bigr]      \Biggr\},
\eea 
where $C_A = N_c$, $\beta_0 = 11N_c/3-2n_f/3$, and $N_c$ is the number of colors. $n_Q~(n_q)$ is a number of heavy- (light-) quark flavors, hence the total number of active quark flavors is given by $n_f = n_q+n_Q$. The functions $h$ and $g$ are
\bea
\label{fh} 
h(b) &=& \int^1_0 dz z(z^2+(1-z)^2) \ln[z^2(1-z)^2+b], \\
\label{fj}
j(b) &=& 2\int^1_0 dz \frac{z^4(1-z)^3}{z^2(1-z)^2+b}.
\eea
For $b=0$ the functions are easily integrated, giving $h(0)=-13/9$ and $j(0)=1/6$. So, as $m_i \to 0$, we easily see that Eq.~(\ref{NLOintgJ}) reproduces the massless result given in Ref.~\cite{Dai:2016hzf}. 

If we combine $\mc{J}_g$ with the out-jet contribution from $g\to g$, similar to Eq.~(\ref{indHQFFJQ}), we can obtain the gluon FFJ for $g\to J_g$.\footnote{\baselineskip 3.0ex
Note that $D_{J_g/g}$ includes the processes $g\to q\bar{q}$ and $g\to Q\bar{Q}$ in the jet, where
$J_g$ is the jet initiated by gluon. 
When the quark pair from the gluon are exactly collinear with each other, we have an IR divergent term or a term sensitive to the  quark mass $\ln (\mu^2/m^2)$. These terms are cancelled by the self-energy interactions of the gluon. 
So $D_{J_g/g}$ is not sensitive to the IR nor the quark mass like other FFJs. 
}
The renormalized result is given as 
\bea
D_{J_g/g}(z;ER',m_i,\mu) 
&=& \delta(1-z) + \frac{\as C_A}{2\pi} \Biggl\{\delta(1-z)\Bigl[\frac{\beta_0}{2C_A}\ln\frac{\mu^2}{p_{J}^{+2}t^2}+\frac{67}{9}-\frac{2\pi^2}{3}-\frac{23}{18}\frac{n_q}{C_A}\nnb\\
&&\hspace{-1.5cm}+\frac{1}{C_A}\sum_{i=1}^{n_Q} \Bigl(h\bigl(\frac{m_i^2}{p_J^{+2}t^2}\bigr) +j\bigl(\frac{m_i^2}{p_J^{+2}t^2}\bigr) \Bigr)\Bigr]
+2\ln\frac{\mu^2}{q_+^2t^2}\Bigl[\frac{z}{(1-z)_+}+\frac{1-z}{z}+z(1-z)\Bigr] \nnb\\
\label{Dggnlo}
&&\hspace{-1.5cm}-4\Bigl[\frac{z\ln z}{(1-z)_+}+z\Bigl(\frac{\ln(1-z)}{1-z}\Bigr)_+ + \ln [z(1-z)]\Bigl(\frac{1-z}{z}+z(1-z)\Bigr)\Bigr]\Biggr\},
\eea 
where $q_+$ is the mother parton's momentum, which can be given by $p_J^+/z$. Also from Eq.~(\ref{MgQout}) we obtain the gluon FFJ for $g\to J_Q$ process 
\bea
\label{DgQnlo}
D_{J_Q/g}(z;ER',m,\mu)  &=& \frac{\as}{2\pi} \Biggl\{\frac{z^2+(1-z)^2}{2}
\ln \frac{\mu^2}{z^2(1-z)^2 q_+^2t^2+m^2} \\
&&~~~~~~-z(1-z) \frac{z^2(1-z)^2 q_+^2t^2}{z^2(1-z)^2 q_+^2t^2+m^2}\Biggr\}.\nnb
\eea
Like the heavy quark case, the gluon initiated processes satisfy the momentum sum rule,
\be
\label{sumr} 
\sum_i \int^1_0 dz z D_{J_i/g}(z) = \int^1_0 dz z \Bigl(D_{J_g/g}(z)+2n_q D_{J_q/g}(z)+2n_Q D_{J_Q/g}(z)\Bigr)=1.
\ee
Also, as can be seen in Eqs.~(\ref{Dggnlo}) and (\ref{DgQnlo}), the heavy-quark mass does not affect the renormalization behavior, which still follows DGLAP evolution.  

\section{Heavy Quark Mass Effects on Jet Fragmentation} 
\label{HQFFJ}

In this section we consider the $p_T$ spectrum of a subjet or hadron inside an observed jet with a small radius, where $p_T$ is the  momentum transverse to the beam axis. The relevant formalism using the FFJs was introduced in Ref.~\cite{Dai:2016hzf}. This formalism can be extended to the situation  when heavy quarks are involved in the jet,
\be
\label{factJFF} 
\frac{d\sigma}{dydp_T^Jdz} = \sum_{i,k=q(\bar{q}),Q(\bar{Q}),g} \int^1_{x_J} \frac{dx}{x} \frac{d\sigma_{i}(y,x_J/x;\mu)}{dydp_T^i} D_{J_k/i}(x,\mu) D_{A/J_k} (z),
\ee
where $\sigma_i$ is the scattering cross section to the parton $i$, and $k$ is an initial 
(and primary) parton for the jet $J_k$. This factorization above between the FFJ and the JFF holds to order $\alpha_s$. $A=j,H$ represents a subjet $(j)$ or a hadron $(H)$ that is observable inside the jet. Because we are interested in the high-$p_T$ region, the rapidity $y\lesssim \mc{O}(1)$ is small. The momentum fraction variables are $x_J = p_T^J/Q_T$, $x=p_T^J/p_T^i$, and $z=p_T^A/p_T^J$, where $p_T^x$ is the transverse momentum of object $x$ and $Q_T$ is the maximal jet transverse momentum for the given rapidity $y$. 
 
In Eq.~(\ref{factJFF}) the JFF, $D_{A/J_k}(z)$, describes the fragmenting processes from a jet to a jet containing $A$ and represents the probability that a final subjet or hadron has momentum fraction $z$ of the total jet momentum. The JFFs are normalized to satisfy the momentum sum rules
\be
\label{msJFF}
\sum_H \int^1_0 dz z D_{H/J_k}(z) = 1,~~~\sum_l \int^1_0 dz z D_{j_l/J_k}(z) = 1,
\ee
where $l$ denotes the initial parton for the subjet. 

As was implicitly shown in Eq.~(\ref{factJFF}), the JFFs are independent of the renormalization scale (except for the dependence in the coupling $\as$). Since the FFJs $D_{J_k/i}$ follow DGLAP evolution, the convolution of the FFJs and $d\sigma_i/(dydp_T)$ is scale invariant. However, the JFFs can be governed by two distinct scales, $\mu_J$ and $\mu_A$, where $\mu_J\sim E_JR'$ is a typical scale for the jet and $\mu_A$ is the typical scale for the subjet or the hadron. Therefore the JFFs can be further  factorized 
\be
\label{JFFfact} 
D_{A/J_k}(z) = \int^1_z \frac{dw}{w} K_{l/k}(z/w,\mu) D_{A/l}(w,\mu),
\ee
where $K_{l/k}$ are the splitting kernels inside the jet $J_k$ that can be perturbatively calculated at the scale $\sim E_JR'$ and $D_{A/l}$ are the fragmentation functions to $A=j,H$ to be evaluated at the lower scale. 
Since $K_{l/k}$ is independent of the final state $A$, we can easily reconstruct the perturbative amplitudes for the various JFFs once we obtain the perturbative results of $D_{A/l}$.

If we consider the partonic level JFFs in Eq.~(\ref{JFFfact}), we need to employ parton fragmentation functions on the right-hand side of the equation. With a heavy quark as one of the partons, the HQFFs at NLO are
\bea
\label{HFQQ}
D_{Q/Q}(z,\mu) &=& \delta(1-z)+\frac{\as C_F}{2\pi} \Biggl[\frac{1+z^2}{1-z}\Bigl(\ln\frac{\mu^2}{m^2(1-z)^2}-1\Bigr)\Biggr]_{+},\\
\label{HFgQ}
D_{g/Q} (z;m,\mu) &=& \frac{\as C_F}{2\pi} \frac{1+(1-z)^2}{z} \Bigl(\ln\frac{\mu^2}{m^2(1-z)^2}-1\Bigr),\\
\label{HFQg}
D_{Q/g} (z;m,\mu) &=& \frac{\as C_F}{2\pi} \frac{z^2+(1-z)^2}{2} \ln\frac{\mu^2}{m^2}.
\eea 
The fragmentation for $g\to g$ at NLO also depends on the heavy quark masses due to the gluon self-energy interactions. The bare fragmentation function can be written as  
\bea
D_{g/g} (z;m_i,\mu) &=& \delta(1-z) + \frac{\as}{4\pi}\Biggl\{4C_A\Bigl(\frac{1}{\veps_{\mr{UV}}} -\frac{1}{\veps_{\mr{IR}}}\Bigr)\Bigl[\frac{z}{(1-z)_+}+\frac{1-z}{z}+z(1-z)\Bigr]\nnb\\
\label{Dgg} 
&&~~~~~+\delta(1-z)\Bigl[\beta_0 \frac{1}{\veps_{\mr{UV}}} - \Bigl(\frac{11}{3}N_c-\frac{2}{3}n_q\Bigr)\frac{1}{\veps_{\mr{IR}}}
-\frac{2}{3}\sum_{i}^{n_Q} \ln \frac{\mu^2}{m_i^2}\Bigr]\Biggr\}.
\eea

On the right-hand side of Eq.~(\ref{JFFfact}) we can also consider the fragmentation functions to a subjet (FFsJs), $D_{j_l/k}$. 
If we define the subjet with a subjet merging condition $\theta < r'$  similar to Eq.~(\ref{jmerging}), 
the FFsJs share the same definitions as the FFJs shown in Eqs.~(\ref{defHQFFJ}) and (\ref{gJFFpar}). In this case the only difference is that $ER'$ in the argument of the FFJs should be changed to $E_J r'$ for the FFsJs.
 
Therefore, if we consider the JFFs for the heavy hadron (subjet) in the limit $E_JR'\gg m$ ($R\gg r$), the perturbative results at the fixed order in $\as$ involve large logarithms due to the large scale difference. In this case, using the factorization theorem shown in Eq.~(\ref{JFFfact}), we can systematically resum the large logarithms through RG evolutions of $K_{l/k}$ and $D_{A/l}$.

We can read off the renormalization behavior for the perturbative kernels from Eq.~(\ref{JFFfact}), since the fragmentation functions on the right-hand side follow DGLAP evolutions and the JFFs on the left-hand side are scale invariant. Hence the RG equations for $K_{l/k}$ are simply  
\be\label{DGLAPK} 
\frac{d}{d\ln\mu} K_{l/k} (x,\mu) = -\frac{\as(\mu)}{\pi}\int_x^1 \frac{dz}{z} P_{l'k}(z) K_{l/l'} (x/z,\mu),  
\ee
where $P_{l'k}$ are the DGLAP kernels 
\bea 
\label{pqq}
P_{qq}(z) &=& C_F \Bigl[\frac{3}{2}\delta(1-z) + \frac{1+z^2}{(1-z)_+} \Bigr], \\
\label{pgq}
P_{gq}(z) &=& C_F \Bigl[\frac{1+(1-z)^2}{z}\Bigr], \\
\label{pqg}
P_{qg}(z) &=& \frac{1}{2} \Bigl[z^2+(1-z)^2], \\
\label{pgg}
P_{gg}(z) &=& \frac{\beta_0}{2}\delta(1-z)+2C_A\Bigl[\frac{z}{(1-z)_+}+\frac{1-z}{z}+z(1-z)\Bigr]. 
\eea

For massless quarks, NLO results for the perturbative kernels were computed in Ref.~\cite{Dai:2016hzf}. Include a heavy quark mass, the perturbative kernels at NLO in $\as$ are
\bea
K_{Q/Q}(z;E_JR',m,\mu)  &=& \delta(1-z) - \frac{\as C_F}{2\pi}\Biggl\{\delta(1-z) \Bigl[f\bigl(\frac{m^2}{p_J^{+2} t^2}\bigr)+g\bigl(\frac{m^2}{p_J^{+2} t^2}\bigr)\Bigr] \nnb \\
&&+\Bigl(\frac{1+z^2}{1-z}\Bigr)_+ \ln\frac{\mu^2}{z^2 p_J^{+2} t^2+m^2}-\Bigl(2\frac{1+z^2}{1-z}\ln(1-z)+1-z\Bigr)_+ \nnb\\
\label{KQQ}
&&-\Bigl(\frac{2z}{1-z}\Bigr)_+\frac{m^2}{z^2 p_J^{+2} t^2+m^2} \Biggr\},\\
K_{g/Q}(z;E_JR',m,\mu)  &=& -\frac{\as C_F}{2\pi}\Biggl\{\frac{1+(1-z)^2}{z}\Bigl(\ln\frac{\mu^2}{(1-z)^2 p_J^{+2} t^2+m^2}-2\ln z\Bigr)-z\nnb\\
\label{KgQ}
&&-\frac{2(1-z)}{z} \frac{m^2}{(1-z)^2 p_J^{+2} t^2+m^2}\Biggr\},
\eea
\bea
K_{g/g}(z;E_JR',m_i,\mu) 
&=& \delta(1-z) - \frac{\as C_A}{2\pi} \Biggl\{\delta(1-z)\Bigl[\frac{\beta_0}{2C_A}\ln\frac{\mu^2}{p_{J}^{+2}t^2}+\frac{67}{9}-\frac{2\pi^2}{3}-\frac{23}{18}\frac{n_q}{C_A}\nnb\\
&&\hspace{-1.5cm}+\frac{1}{C_A}\sum_{i=1}^{n_Q} \Bigl(h\bigl(\frac{m_i^2}{p_J^{+2}t^2}\bigr) +j\bigl(\frac{m_i^2}{p_J^{+2}t^2}\bigr) \Bigr)\Bigr]
+2\ln\frac{\mu^2}{p_J^{+2}t^2}\Bigl[\frac{z}{(1-z)_+}+\frac{1-z}{z}+z(1-z)\Bigr] \nnb\\
\label{Kgg}
&&\hspace{-1.5cm}-4\Bigl[\frac{z\ln z}{(1-z)_+}+z\Bigl(\frac{\ln(1-z)}{1-z}\Bigr)_+ + \ln [z(1-z)]\Bigl(\frac{1-z}{z}+z(1-z)\Bigr)\Bigr]\Biggr\},\\
\label{KQg}
K_{Q/g}(z;E_JR',m,\mu)  &=& -\frac{\as}{2\pi} \Biggl\{\frac{z^2+(1-z)^2}{2}
\ln \frac{\mu^2}{z^2(1-z)^2 p_J^{+2} t^2+m^2} \\
&&~~~~~~~~-z(1-z) \frac{z^2(1-z)^2 p_J^{+2} t^2}{z^2(1-z)^2 p_J^{+2}t^2+m^2}\Biggr\},\nnb
\eea
where $p_J^+ t\sim E_JR'$.  
Note that the heavy quark mass does not affect the renormalization behaviors of $K_{l/k}$ similar to the case for FFJs. 
In the limit $E_JR'\gg m$ we can safely ignore the quark mass and these kernels reduce to the massless results obtained in Ref.~\cite{Dai:2016hzf}. 
However, when $E_JR'$ is comparable with $m$ or when the corrections $m/(E_JR')$ give significant enough corrections to be interesting, the above complete results will be useful.  

From the NLO results, we can easily check the momentum sum rule
\be
\label{msK}
\sum_l \int^1_0 dz z K_{l/k}(z) =1.
\ee 
Further, as in Ref.~\cite{Dai:2016hzf}, we have  the relations between the perturbative kernels and FFJs
\be
\label{relDK}
D_{J_k/i}^{(1)} (z;ER',\mu) = -K_{k/i}^{(1)} (z;ER',\mu). 
\ee
Here the superscript represents the relation is true at one loop order and $E_JR'$ in the perturbative kernels has been replaced with $ER'$, where $E$ is the energy of the mother parton. Note that the relations in Eq.~(\ref{relDK}) are still valid when we include the heavy quark mass. Comparing Eqs.~\eqref{KQQ}-\eqref{KQg} with the NLO results of the FFJs with heavy quarks in Sec.~\ref{HQFFJsec}, we clearly see these relations hold.

\section{Inclusive $b$-jet production}
\label{IbJp}

Inclusive $b$-jet production is a good arena for studying perturbative QCD and probing Standard Model predictions, since hadronic uncertainty should be negligible compared to $b$-hadron production~\cite{Frixione:1996nh}. Recently the $b$-jet production rate in $pp$ collision has been measured with a jet radius $R=0.5$~\cite{Chatrchyan:2012dk}.\footnote{It has been observed that the small $R$ approximation for an inclusive jet process works well even up to $R \lesssim 0.7$~\cite{Jager:2004jh}. We thus believe that the CMS experiment with $R=0.5$~\cite{Chatrchyan:2012dk} can be legitimately compared with our analysis using the small $R$ approximation, which will be performed in Sec.~\ref{numeric}.
Dominant finite size effects of $\mc{O}(R^2)$ need to be considered for more a precise estimation. This is beyond the scope of this paper.}
Based on this result, the production in heavy ion collision has been analyzed to examine heavy quark jet quenching~\cite{Huang:2013vaa}. 

In this section, we consider inclusive $b$-jet production in the regime $E_J \gg E_JR' \gg m_b$.
Here the $b$-jet is defined to contain one or more $b(\bar{b})$-quarks inside the jet. As is well known, usual jet production is insensitive to long-distance interactions if we employ an IR safe jet algorithm and the jet scale $E_JR'$ is much larger than the long-distance scale. Thus we might naively speculate that the $b$-jet production in the limit $E_JR' \gg m_b$ would be insensitive to the $b$-quark mass, and hence we might be able to take the limit $m_b \to 0$. However, as pointed out in Ref.~\cite{Banfi:2006hf,Banfi:2007gu}, the $b$-jet is actually quite sensitive to the heavy quark mass. As a gluon splits into a $b\bar{b}$ pair with zero angle inside the jet, the amplitude becomes singular as $m_b$ goes to zero. 

For inclusive $b$-jet production in the limit $E_J \gg E_JR' \gg m_b$, this sensitivity appears as a term with $\ln E_JR'/m_b$ in the jet initiated by a gluon at order $\as$. So, for reliable perturbative predictions, these large logarithms need to be resummed to all orders in $\as$. 
In order to do this, we can employ the gluon to $b$-quark JFF, $D_{b/J_{g}}$, which describes the splitting process of $g\to b\bar{b}$ inside a jet. As $E_JR' \gg m_b$, this JFF can be further factorized into the splitting kernels $K_{l/g}$ and the $b$-quark fragmentation functions $D_{b/l}$ as illustrated in Eq.~\eqref{JFFfact}. Then, through RG evolution of each factorized part, we can consistently resum the large logarithms of $E_JR'/m_b$. In this section we will describe the inclusive $b$-jet production using the FFJs and present the procedure of resumming the large logarithms in $g\to b\bar{b}$ process using the factorization theorem for the JFF $D_{b/J_{g}}$.


\subsection{Analysis using the fragmentation functions to a $b$-jet}
\label{sect-ffbj}
Using the FFJs, the inclusive $b$-jet production at NLO can be written as 
\bea
\frac{d\sigma}{dyp_T^J} &=&\int^1_{x_J} \frac{dx}{x} \Biggl\{\frac{d\sigma_{b}(y,x_J/x;\mu)}{dydp_T} D_{J_b/b}(x;E_JR',m_b,\mu) +\frac{d\sigma_{\bar{b}}(y,x_J/x;\mu)}{dydp_T} D_{J_{\bar{b}}/\bar{b}}(x;E_JR',m_b,\mu)\nnb \\
\label{bjetnv}
&&+\frac{d\sigma_{g}(y,x_J/x;\mu)}{dydp_T}\Bigl[2 D_{J_b/g}(x;E_JR',m_b,\mu) +\delta(1-x)\cdot \mc{M}^{\mr{In}}_{g\to b\bar{b}}(E_JR',m_b)\Bigr]\Biggr\},
\eea
where $E_JR'=p_T^JR$ at a hadron collider, and we identified separately the fragmenting processes $g\to b$ and $g\to \bar{b}$. 
Here $\mc{M}^{\mr{In}}_{g\to b\bar{b}}$ is the amplitude squared for $g\to b\bar{b}$ inside the jet. So the term $\delta(1-x) \mc{M}^{\mr{In}}_{g\to b\bar{b}}$ is the contribution of $g\to b\bar{b}$ to the gluon FFJ. 

In order to describe the $b$-jet production in a straightforward way, we introduce the fragmentation functions to the $b$-jet (FFbJs), $D_{\mathsf{J_b}/i} (z)$, where $\msf{J_b}$ represents the $b$-jet that includes at least one $b(\bar{b})$-quark.\footnote{
In our convention, $J_b$ represents a jet {\it initiated} by a $b$-quark, while $\msf{J_b}$ represents the (physical) $b$-jet that contains at least one $b(\bar{b})$. 
$J_b$ and $\msf{J_b}$ become different at NLO in $\as$. For example, when a gluon initiates a jet and splits into a $b\bar{b}$ pair inside the jet, it contributes to $\msf{J_b}$ as seen in Eqs.~\eqref{bjetnv} and \eqref{inb}.
}  
Then the scattering cross section in Eq.~(\ref{bjetnv}) can be rewritten as 
\bea
\frac{d\sigma}{dyp_T^{J}} &=&\int^1_{x_J} \frac{dx}{x} \Biggl\{\frac{d\sigma_{b}(y,x_J/x;\mu)}{dydp_T} D_{\msf{J_b}/s}(x;E_JR',m_b,\mu) \nnb \\ 
\label{inb}
&&+\frac{d\sigma_{g}(y,x_J/x;\mu)}{dydp_T}D_{\msf{J_b}/g}(x;E_JR',m_b,\mu)\Biggr\},
\eea
where $D_{\msf{J_b}/s}$ is the singlet FFbJ defined as 
$D_{\msf{J_b}/s}=D_{\msf{J_b}/b}+D_{\msf{J_b}/\bar{b}}$. We are using $d\sigma_b/(dydp_T) = d\sigma_{\bar{b}}/(dydp_T)$, ignoring the charge asymmetry. 
We also have suppressed the light quark contributions to the $b$-jet, i.e., $D_{\msf{J_b}/q(\bar{q})}$, which first appear at two loops. 
From Eq.~(\ref{bjetnv}), $D_{\msf{J_b}/g}$ at the first order in $\as$ is
\be
\label{gjbnlo}
D_{\msf{J_b}/g} (x;E_JR',m_b,\mu) = 2 D_{J_b/g}(x,E_JR',m_b,\mu) +\delta(1-x)\cdot \mc{M}^{\mr{In}}_{g\to b\bar{b}}(E_JR',m_b). 
\ee

In Eq.~(\ref{MgQin}), we calculated $g\to Q\bar{Q}$ inside a jet with the heavy quark momentum fraction $z$ specified. The result $\mc{M}_{g\to Q}^{R,\mr{In}}(z)$ in Eq.~(\ref{MgQin}) can also be considered as the leading result of the JFF $D_{Q/J_g}(z)$. Therefore using the result in Eq.~(\ref{MgQin}), we obtain $\mc{M}^{\mr{In}}_{g\to b\bar{b}}$:
\bea
\mc{M}^{\mr{In}}_{g\to b\bar{b}} (E_JR',m_b) &=& 2 \int^1_0 dz z D_{b/J_g}(z;E_JR',m_b) =2 \int^1_0 dz z \mc{M}_{g\to Q}^{R,\mr{In}}(z) \nnb \\
\label{Mgbbfix}
&=& \frac{\as}{2\pi} \Bigl[\frac{1}{3} \ln \frac{p_J^{+2}t^2}{m_b^2}+h\bigl(\frac{m_b^2}{p_J^{+2}t^2}\bigr) +j\bigl(\frac{m_b^2}{p_J^{+2}t^2}\Bigr)\Bigr],
\eea
where the functions $h$ and $j$ are defined in Eqs.~(\ref{fh}) and (\ref{fj}).  
The presence of the term with $\ln (p_J^{+2}t^2)/m_b^2$ gives a large uncertainty for the fixed order result in $\as$, and we have to resum these large logarithms to all orders in $\as$ for a reliable prediction. 
For $E_JR'\gg m_b$, the logarithmic term at order $\as$ in Eq.~(\ref{Mgbbfix}) and its resummed result can be estimated to be $\mO(1)$, which implies that the gluon fragmentation $D_{\msf{J_b}/g}$ is not suppressed when compared with $D_{\msf{J_b}/s}$.

In Eq.~(\ref{inb}), if we choose the factorization scale $\mu_F \sim p_T^J$, the resummation of large logarithms with small $R$ is crucial. The resummation can be performed by RG evolutions of the FFbJs from $\mu\sim E_JR'$ to $\mu_F\sim p_T^J$, which should be equivalent to DGLAP evolution.  The RG equations are
\be
\label{RGEb}
\frac{d}{d\ln\mu}\binom{D_{\msf{J_b}/s}(x;\mu)}{D_{\msf{J_b}/g}(x;\mu)} = \frac{\as}{\pi} \int^1_{x}\frac{dz}{z}\begin{pmatrix} P_{qq}(z) & 2P_{gq}(z) \\ P_{qg}(z) & P_{gg}(z) \end{pmatrix} 
\binom{D_{\msf{J_b}/s}\bigl(\frac{\displaystyle x}{\displaystyle z};\mu\bigr)}{D_{\msf{J_b}/g}\bigl(\frac{\displaystyle x}{\displaystyle z};\mu\bigr)}.
\ee
Note that, we have included a factor of  2 in front of $P_{gq}$ for the RG equation of the singlet FFbJ in Eq.~\eqref{RGEb}. This is necessary since the singlet FFbJ is defined as
$D_{\msf{J_b}/s}=D_{\msf{J_b}/b}+D_{\msf{J_b}/\bar{b}}$, and each of 
$D_{\msf{J_b}/b(\bar{b})}$ satisfy the following RG equation
\be
\frac{d}{d\ln\mu} D_{\msf{J_b}/k}(x) =\frac{\as}{\pi} \int^1_{x}\frac{dz}{z} \Bigl(P_{qq}(z) D_{\msf{J_b}/k}(\frac{x}{z}) +P_{gq}(z) D_{\msf{J_b}/g}(\frac{x}{z})\Bigr),~~~k=b,~\bar{b}.
\ee

After taking the $N$-th moments,
\be
f(N) = \int^1_0 dx x^{-1+N} f(x),
\ee
and solving the RG equations, we obtain the evolved results in moment space 
\be
\label{Dbevo}
\binom{D_{\msf{J_b}/s}(N;\mu_F)}{D_{\msf{J_b}/g}(N;\mu_F)} = \Biggl[\left(\frac{\as(\mu_F)}{\as(\mu_J)}\right)^{-\frac{2\lambda_+}{\beta_0}}M_+ + \left(\frac{\as(\mu_F)}{\as(\mu_J)}\right)^{-\frac{2\lambda_-}{\beta_0}}M_-\Biggr]
\binom{D_{\msf{J_b}/s}(N;\mu_J)}{D_{\msf{J_b}/g}(N;\mu_J)}, 
\ee
where the scales are roughly $\mu_F \sim p_T^J$ and $\mu_J \sim E_JR'$.  $\lambda_{\pm}$ is 
\be
\label{eigen}
\lambda_{\pm} = \frac{1}{2}\Bigl[P_{qq}(N)+P_{gg}(N)\pm \sqrt{(P_{qq}(N)-P_{gg}(N))^2+8P_{gq}(N)P_{qg}(N)}\Bigr],
\ee
and the matrices $M_{\pm}$ are
\be
M_{\pm}=\frac{1}{\lambda_{\pm}-\lambda_{\mp}}\begin{pmatrix} P_{qq}(N) -\lambda_{\mp} & 2P_{gq}(N) \\ P_{qg}(N) & P_{gg}(N)-\lambda_{\mp} \end{pmatrix}. 
\ee
Although $D_{\msf{J_b}/g}(N;\mu_J)$ in Eq.~(\ref{Dbevo}) starts at the order $\as$, it
can be power-counted as $\mc{O}(1)$, similar to $D_{\msf{J_b}/b}$, due to the large logarithmic term $\ln (E_JR'/m_b)$.
Therefore, even at leading logarithm (LL) accuracy, we must keep a nonzero $D_{\msf{J_b}/g}(N;\mu_J)$.

\subsection{Resummation of large logarithms in the $g\to b\bar{b}$ process}
\label{Rgbb}

As seen in Eq.~(\ref{Mgbbfix}), since $\mc{M}^{\mr{In}}_{g\to b\bar{b}}$ can be expressed as the integral of $D_{b/J_g}$,  the resummation of the large logarithms can be accomplished using the factorization formula for the JFF. Using  Eq.~(\ref{JFFfact}), we write $\mc{M}^{\mr{In}}_{g\to b\bar{b}}$ 
\bea\label{Mgbbfac}
\mc{M}^{\mr{In}}_{g\to b\bar{b}} (E_JR',m_b) &=& 2 \int^1_0 dz z D_{b/J_g}(z;E_JR',m_b)\\
&=&  2\sum_{l=g,b} \bar{K}_{l/g}(E_JR',m_b,\mu_F)\cdot\bar{D}_{b/l}(m_b,\mu_F),
\nnb
\eea
where the functions $\bar{f}$ represent 
\be
\bar{f} = \int^1_0 dz z f(z).
\ee
The factorization scale $\mu_F$ can be chosen arbitrarily.  The scale to minimize the large logarithms in $\bar{K}_{l/g}~(\bar{D}_{b/l})$ is $\mu \sim E_JR'~(m_b)$. Therefore, if we choose $\mu_F \sim E_JR'$, we have to perform RG evolution from $\mu_F$ to $m_b$ for $\bar{D}_{b/l}$. For $\mu_F \sim m_b$, RG evolution between $\mu_F$ and $E_JR'$ is required for a reliable result of $\bar{K}_{l/g}$. Through these RG evolutions we can  resum the large logarithmic terms  $\ln E_JR'/m_b$.

We set $\mu_F \sim E_JR'$ and evolve $\bar{D}_{b/l}$ from $\mu_F$ to $m_b$ at LL. Since the HQFFs, $D_{b/l}$, follow DGLAP evolution, the RG equations for $\bar{D}_{b/l}$ are
\be
\label{RGED}
\frac{d}{d\ln\mu}\binom{\bar{D}_{b/b}(\mu)}{\bar{D}_{b/g}(\mu)} = \frac{\as}{\pi} \begin{pmatrix} \bar{P}_{qq} & \bar{P}_{gq} \\ \bar{P}_{qg} & \bar{P}_{gg} \end{pmatrix} 
\binom{\bar{D}_{b/b}(\mu)}{\bar{D}_{b/g}(\mu)}.
\ee
Solving, we obtain
\be
\label{Dblevo}
\binom{\bar{D}_{b/b}(\mu_F)}{\bar{D}_{b/g}(\mu_F)} = \Biggl[\left(\frac{\as(\mu_F)}{\as(\mu_b)}\right)^{-\frac{2\bar{\lambda}_+}{\beta_0}}\bar{M}_+ + \left(\frac{\as(\mu_F)}{\as(\mu_b)}\right)^{-\frac{2\bar{\lambda}_-}{\beta_0}}\bar{M}_-\Biggr]
\binom{\bar{D}_{b/b}(\mu_b)}{\bar{D}_{b/g}(\mu_b)},
\ee
where $\mu_b \sim m_b$. $\bar{\lambda}_{\pm}$ and $\bar{M}_{\pm}$ are, respectively, 
\be
\label{eigenbar}
\bar{\lambda}_{\pm} = \frac{1}{2}\Bigl[\bar{P}_{qq}+\bar{P}_{gg}\pm \sqrt{(\bar{P}_{qq}-\bar{P}_{gg})^2+4\bar{P}_{gq}\bar{P}_{qg}}\Bigr],
\ee
and  
\be
\bar{M}_{\pm}=\frac{1}{\bar{\lambda}_{\pm}-\bar{\lambda}_{\mp}}\begin{pmatrix} \bar{P}_{qq} -\bar{\lambda}_{\mp} & \bar{P}_{gq} \\ \bar{P}_{qg} & \bar{P}_{gg}-\bar{\lambda}_{\mp} \end{pmatrix}. 
\ee

Therefore putting Eq.~(\ref{Dblevo}) into Eq.~(\ref{Mgbbfac}), we obtain the resummed results explicitly. 
If we consider the results at LL running and keeping only the LO fixed order terms in $\as$ (LL+LO), we can  remove the term $2\bar{K}_{b/g}(\mu_F) \bar{D}_{b/b}(\mu_F)$ in Eq.~(\ref{Mgbbfac}), since $\bar{K}_{b/g}(\mu_F)$ is already $\mc{O}(\as)$. Furthermore, we put $\bar{K}_{g/g}(\mu_F)=\bar{D}_{b/b}(\mu_b)=1$ and $\bar{D}_{b/g}(\mu_b)=0$.
As a result we obtain 
\be
\label{MgbbLL}
\mc{M}^{\mr{In,LL+LO}}_{g\to b\bar{b}} (E_JR',m_b) \sim 2 \bar{D}_{b/J_g}(\mu_F) 
 =\frac{2\bar{P}_{qg}}{\bar{\lambda}_{\pm}-\bar{\lambda}_{\mp}} \Biggl[\left(\frac{\as(\mu_F))}{\as(\mu_b)}\right)^{-\frac{2\bar{\lambda}_+}{\beta_0}} - \left(\frac{\as(\mu_F)}{\as(\mu_b)}\right)^{-\frac{2\bar{\lambda}_-}{\beta_0}}\Biggr].
\ee
Expanding the above result in terms of $\as(\mu_F)$ using
\be
\frac{\as(\mu_F)}{\as(\mu_b)} \sim 1-\beta_0 \frac{\as(\mu_F)}{2\pi}\ln\frac{\mu_F}{\mu_b},
\ee
we obtain  
\be
\label{Mgbbas}
\mc{M}^{\mr{In,LL+LO}}_{g\to b\bar{b}}(E_JR',m_b) = \frac{\as}{3\pi} \ln \frac{\mu_F}{\mu_b} + \cdots = \frac{1}{3}\frac{\as}{2\pi} \ln \frac{(E_JR')^2}{m_b^2}+\cdots
\ee
We see that the result in Eq.~(\ref{MgbbLL}) correctly resums the series of large logarithms of $E_JR'/m_b$, which starts with the logarithmic term in Eq.~(\ref{Mgbbfix}). 


\section{Numerical implications of the heavy quark mass}
\label{numeric}
\begin{figure}
\includegraphics[scale=0.75]{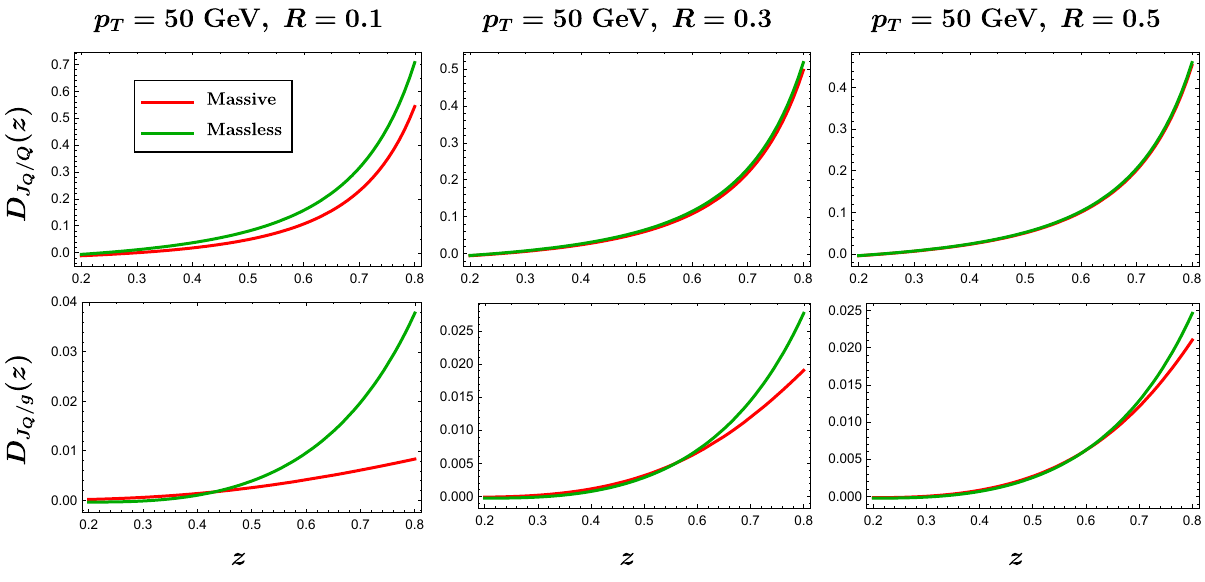}
\caption{Comparison of the heavy quark and massless FFJs at the jet scale $\mu_J \equiv \sqrt{(p_T^{J} R)^2 + m^2}$. The heavy quark mass is taken to be the $b$-quark mass for the heavy FFJs. }
\label{fig-mass-massless}
\end{figure}
In this section we first discuss the heavy quark mass effects on the FFJs/FFbJs and then apply the formalism to $b$-jet production, comparing the results with CMS data~\cite{Chatrchyan:2012dk}.
In Figure~\ref{fig-mass-massless} we have compared heavy quark FFJs ($D_{J_Q/Q}$ and $D_{J_Q/g}$) with the massless case.  
For the heavy quark FFJs, we treat the $b$-quark as the heavy quark ($Q=b$) and ignored the masses of the charm and other light quarks. 
The heavy quark FFJs are evaluated at the jet scale $\mu_J = 
\sqrt{(p_T^{J} R)^2 + m_b^2}$, where $m_b$ is  set to 4.8~GeV. For the massless FFJs, we take the limit $m_b \to 0$ or alternatively use the result of Ref.~\cite{Dai:2016hzf}. It is evident from Fig.~\ref{fig-mass-massless} that the mass effect is significant when the jet scale approaches the heavy quark mass scale. The gluon FFJ has a stronger quark mass dependence than the quark FFJ. Even though $D_{J_Q/Q}$ is larger than $D_{J_Q/g}$ in magnitude (as is shown Fig.~\ref{fig-mass-massless}), the contribution from $D_{J_Q/g}$ can be comparable to $D_{J_Q/Q}$ at the LHC due to the large cross section to gluons. As a result it is important to consider the mass corrections to $D_{J_Q/g}$ when we consider jet production.

\begin{figure}
\includegraphics[scale=0.75]{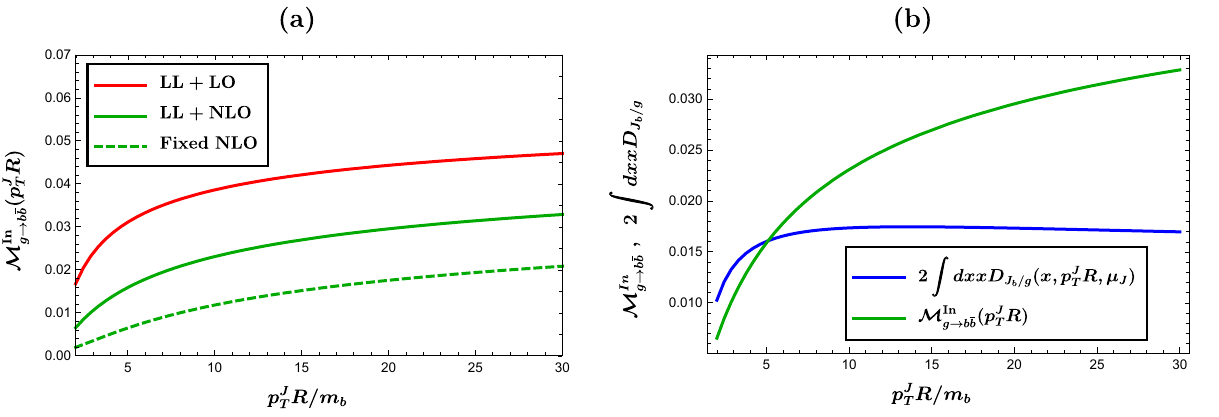}
\caption{Resummation of $\log(m_b)$. Horizontal axis are in units of the $b$ quark mass $m_b$. (a) Solid curves are resummed ${\cal M}_{g\to b\bar{b}}^{\rm In}$ (with the red curve at LO and the green at NLO) and dashed curves are fixed order ${\cal M}_{g\to b\bar{b}}^{\rm In}$ at NLO.  (b) Comparison of the contributions to gluon initiated $b$-jets from $D_{J_b/g}$ (blue) and ${\cal M}_{g\to b\bar{b}}^{\rm In}$ (green). }
\label{fig-log-mQ}
\end{figure}

Next we consider the resummation effects of the logarithms $\ln (p_{T}^J R)/m_b$ in the FFbJ, $D_{\mathsf{J_b}/g} (z)$, where the logarithmic quark mass dependence appears in ${\cal M}_{g\to b\bar{b}}^{\rm In}$, as can be seen in Eq.~(\ref{Mgbbfix}). 
As discussed in Sec.~\ref{Rgbb}, the large logarithms of $(p_T^J R)/m_b$ can be systematically resummed through the factorization formula in Eq.~\eqref{Mgbbfac}.
In Fig.~\ref{fig-log-mQ}-(a), we compared one loop result of ${\cal M}^{\rm In}_{g\to b\bar{b}}$ (``Fixed NLO'') with the resummed results at LL accuarcy (``LL+LO'' and ``LL+NLO'').
Here the result at LL+LO corresponds to Eq.~\eqref{MgbbLL}, and the result at LL+NLO keeps the NLO results of $\bar{K}_{l/g}(\mu_F \sim p_T^JR)$ and $\bar{D}_{b/l}(\mu_b)$ in the factorization formula (Eq.~(\ref{Mgbbfac})) and the resummed formula (Eq.~\eqref{Dblevo}) respectively. The fixed NLO result for ${\cal M}^{\rm In}_{g\to b\bar{b}}$ has been presented in Eq.~\eqref{Mgbbfix}. Both of the resummed results significantly change the fixed NLO result and 
enhance $g\to b\bar{b}$ contribution to $b$-jets roughly by 100\%-200\%.

As shown in Eq.~\eqref{gjbnlo}, there are two type of contributions for the gluon initiated $b$-jet production: $2D_{J_b/g}$ and ${\cal M}_{g\to b\bar{b}}^{\rm In}$. 
To compare the relative size of $2D_{J_b/g}$ with ${\cal M}_{g\to b\bar{b}}^{\rm In}$, we have considered the first moments of $2D_{J_b/g}$, 
\begin{equation}
2\bar{D}_{J_b/g} (\mu_J) \equiv 2\int^1_0 dx x D_{J_b/g}(x,\mu_J).
\end{equation}
In Fig.~\ref{fig-log-mQ}-(b), we show the sensitivity of $2\bar{D}_{J_b/g}$ and ${\cal M}_{g\to b\bar{b}}^{\rm In}$ to $m_b$ by varying $p_T^JR$ in units of $m_b$. 
For ${\cal M}_{g\to b\bar{b}}^{\rm In}$, we used the result at the accuracy of LL+NLO.
As $p_T^JR$ increases, $2\bar{D}_{J_b/g}$ becomes insensitive to the difference between $p_T^JR$ and $m_b$, while ${\cal M}_{g\to b\bar{b}}^{\rm In}$ is still sensitive since ${\cal M}_{g\to b\bar{b}}^{\rm In}$ involves the logarithm of $(p_T^J R)/m_b$. Also we see that ${\cal M}_{g\to b\bar{b}}^{\rm In}$ becomes dominant over $2\bar{D}_{J_b/g}$ as $p_T^J R \gg m_b$. 


\begin{figure}
\includegraphics[scale=0.8]{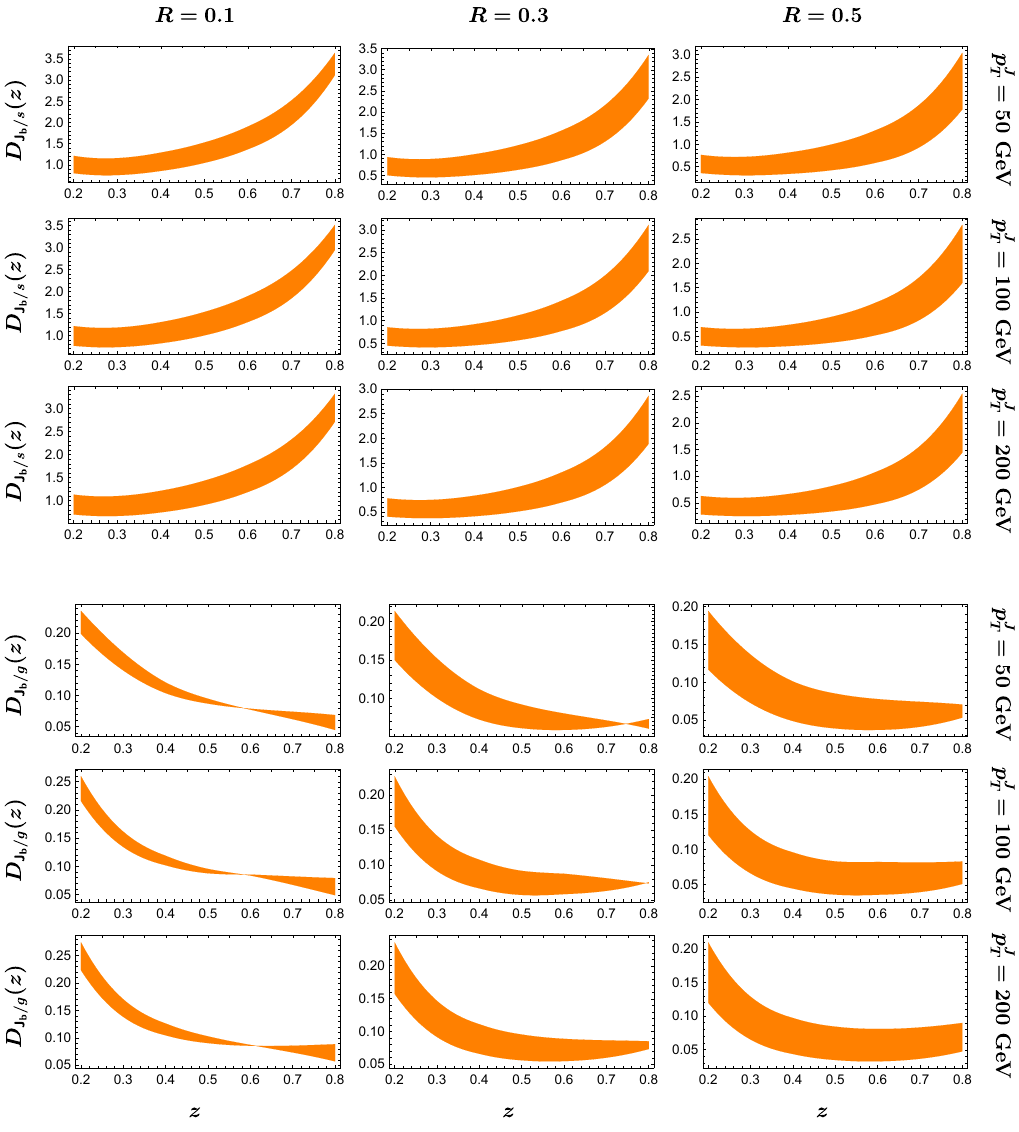}
\caption{FFbJs (defined in Section~\ref{sect-ffbj}) with factorization scale $\mu_F$ equal to initial parton $p_{T,\mathrm{parton}}$. The first three rows are the quark initiated FFbJs  $D_{\msf{J_b}/s}=D_{\msf{J_b}/b}+D_{\msf{J_b}/\bar{b}}$ and the last three rows correspond to the gluon initiated FFbJs. The error bands are obtained by varying the jet scale from $\frac{1}{2}(m_b+\sqrt{(p_T^J R)^2 + m_b^2})$ to $2\sqrt{(p_T^J R)^2 + m_b^2}$.}
\label{fig-FFbJ}
\end{figure}

As an application of FFJs, we consider the inclusive $b$-jet production at the LHC. To study inclusive $b$-jet production, we need to employ the FFbJs (defined in Section~\ref{sect-ffbj}), which describe the production of a jet containing at least one $b$ quark. In Fig.~\ref{fig-FFbJ}, we show $b$-quark and gluon initiated FFbJs at the factorization scale $\mu_F$ equal to initial parton $p_{T,\mathrm{parton}}$, i.e., we solve DGLAP equations Eq.~(\ref{RGEb}) to evolve FFbJs from jet scale $\sqrt{(p_T^J R)^2 + m_b^2}$ to initial parton $p_{T,\mathrm{parton}}$. The error bands in the figure are obtained by varying the jet scale from $\frac{1}{2}(m_b+\sqrt{(p_T^J R)^2 + m_b^2})$ to $2\sqrt{(p_T^J R)^2 + m_b^2}$. Note that we choose $\frac{1}{2}(m_b+\sqrt{(p_T^J R)^2 + m_b^2})$ instead of $\frac{1}{2}\sqrt{(p_T^J R)^2 + m_b^2}$ to make sure that the jet scale chosen is large than $m_b$. 

\begin{figure}
\includegraphics[scale=1.2]{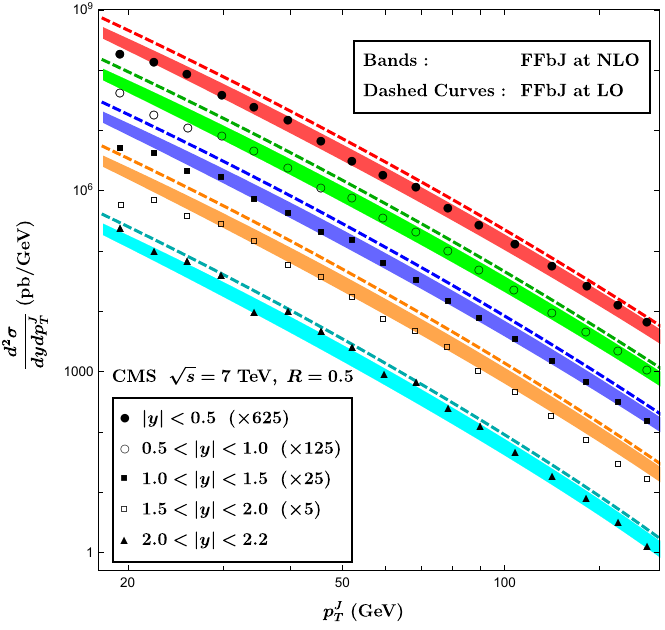}
\caption{$b$-jet production at the LHC \cite{Chatrchyan:2012dk}. The bands correspond to the LO partonic cross section combined with NLO FFbJs and the dashed curves correspond to LO partonic cross section combined with LO FFbJs.}
\label{fig-CMS-Comparison}
\end{figure}

In Fig.~\ref{fig-CMS-Comparison} we show LO and NLO calculations of inclusive $b$-jet production at the LHC based on Eq.~(\ref{bjetnv}) and their comparison with CMS data from Ref.~\cite{Chatrchyan:2012dk}. LO and 
NLO in Fig.~\ref{fig-CMS-Comparison} only refer to the calculations of FFbJs, since we  use LO partonic cross sections (with PDF sets from CTEQ6L1) to calculate $b$ and gluon production rates.  
All the FFbJs are evolved from jet scale $\mu_J=\sqrt{(p_T^JR)^2+m_b^2}$ to $p_{T,\mathrm{parton}}$ of the initiating partons (gluons and $b$'s), and we choose $p_{T,\mathrm{parton}}$ to be the factorization scale. 
In Fig.~\ref{fig-CMS-Comparison}, the NLO calculations look to be more consistent with the data than the LO results. 
The method of obtaining the error bands is the same as that of Fig.~\ref{fig-FFbJ}.  Note that the error estimation only comes from scale variation of FFbJs. Again, the ``NLO calculation" shown in Fig.~\ref{fig-CMS-Comparison} is only a partial calculation of the full NLO computation, since we are using LO partonic cross sections and the NLO partonic cross sections will also modify the normalization of the cross sections.
 To get a better estimation, we need the full NLO calculation as well as the resummed results of $\ln(1-z)$ as $z\to 1$ in the FFJs~\cite{Dai:2017dpc,Liu:2017pbb,Liu:2018ktv}. We leave a more precise analysis and the studies of many other interesting phenomenology (such as top jets and Higgs decays to heavy quark jets) to future work.

\section{Conclusion}
\label{concl}

We studied the process of a parton fragmenting into a heavy-quark jet, keeping the heavy-quark mass nonzero in the FFJ originally introduced in Ref.~\cite{Dai:2016hzf}.  When the typical jet scale is not too large compared to the quark mass, numerically relevant contributions to the jet cross section can occur. To show this, we first calculated the FFJs to NLO with a nonzero quark mass. These results smoothly reduce to the massless FFJs when taking $m\to0$. We show, not surprisingly, that the FFJs still evolve following the usual DGLAP evolution and that this can be used to write resummed results.  Using this, we are able to show that there are indeed non-negligible numerical corrections, especially when the jet scale is not too large compared to the quark mass.

We then investigated inclusive heavy quark jet fragmentation, using the formula, Eq.~(\ref{factJFF}), also originally introduced in Ref.~\cite{Dai:2016hzf}. This  formula describes the inclusive jet rate as the convolutions of a hard cross section producing an outgoing parton with the FFJ and the JFF. This JFF can be further factored into a perturbative kernel and a fragmentation function. Including a nonzero mass, we calculate these kernels, again showing that they reduce to the massless case when $m\to0$.    

Of particular importance for the fragmentation to a $b$-jet is the contribution from $g\to b\bar{b}$, where both $b$ and $\bar b$ end up inside the jet. This contribution is encoded in what we called $\mc{M}^{\mr{In}}_{g\to b\bar{b}}(E_JR',m_b)$. The logarithmic dependence on the heavy quark mass appears in $\mc{M}^{\mr{In}}_{g\to b\bar{b}}$. We show that $\mc{M}^{\mr{In}}_{g\to b\bar{b}}$ can be written as the integral over $D_{b/J_g}$. Using the factorized result of $D_{b/J_g}$ shown in Eq.~(\ref{Mgbbfac}), we can resum the large logarithms $\ln(E_JR')/m_b$  by running the JFF from $E_J R'$ to $m_b$. The resummation of these large logarithms changes $\mc{M}^{\mr{In}}_{g\to b\bar{b}}$ by order one and must be included to obtain reliable results. We further show that the contribution from  $\mc{M}^{\mr{In}}_{g\to b\bar{b}}$ is numerically as important as the direct fragmentation of a gluon to a $b$-quark, where the $\bar b$ is outside the jet, described by $D_{J_b/g}$.

As an application, we combine the above to study inclusive $b$-jet production at the LHC, which has been measured by the CMS collaboration \cite{Chatrchyan:2012dk}. At lowest order, the theoretical prediction is consistently above the measured rate. Including the NLO contributions to the FFJ (keeping the partonic cross sections LO) reduces the calculated result to agree with the measured rate. This result shows the utility of the FFJs in calculating inclusive jet rates at high-energy colliders. There are a number of future directions where FFJs could be useful, including for instance top jets or Higgs decays to heavy quarks, which we leave to future work.

\acknowledgments

The authors would like to thank Geoffrey Bodwin for helpful discussions on numerical methods while LD was visiting Argonne National Laboratory. 
CK was supported by Basic Science Research Program through the National Research Foundation of Korea (NRF) funded by the Ministry of Science and ICT (Grants No. NRF-2014R1A2A1A11052687, No. NRF-2017R1A2B4010511). AL and LD were supported in part by NSF grant PHY-1519175. This research was also supported by the Munich Institute for Astro- and Particle Physics (MIAPP) of the DFG cluster of excellence ``Origin and Structure of the Universe".



\bibliographystyle{JHEP1}
\bibliography{Jet}


\end{document}